\def\bee{\begin{equation}}
\def\ee{\end{equation}}
\def\co2{CO$_2$}
\def\CO2{CO$_2$}
\def\ch4{CH$_4$}
\def\CH4{CH$_4$}
\def\H2{H$_2$}
\def\h2{H$_2$}
\RequirePackage{fix-cm}
\documentclass[twocolumn]{svjour3}          
\smartqed  
\usepackage{graphicx}
\begin{document}

\title{ Selective adsorption of Carbon Dioxide from Mixed Vapors by Blockage of
  Methane in Graphene Nanoribbons
}

\titlerunning{Selective Adsorption of Carbon Dioxide  in Graphene Nanoribbons}        

\author{Hind Aljaddani \and Silvina M. Gatica$^*$}%


\institute{H. Aljaddani  \at
             Department of Physics and  Astronomy, Howard University, 2355 Sixth St NW, Washington, DC 20059, USA\\             
              \email{hind.aljaddani@howard.edu}           
           \and
           S. Gatica \at
                        Department of Physics and  Astronomy, Howard
                        University, 2355 Sixth St NW, Washington, DC
                        20059, USA\\
                        \email{sgatica@howard.edu}            
}

\date{Received: date / Accepted: date}

\maketitle

\begin{abstract}
We study numerically the adsorption of a mixture of    CO$_2$ and CH$_4$ on a graphite substrate covered by graphene nanoribbons (NRs). The NRs are flat and parallel to the graphite surface, at a variable distance ranging from 6 \AA\ to 14 \AA.    We show that the NRs-graphite substrate acts as an effective  filter for \co2. 
Our study is based on   Molecular Dynamics (MD) simulations.  Methane is considered  a spherical molecule, and  carbon dioxide  is represented as a linear rigid body. 
Graphite is modeled  as a continuous material, while the NRs are  approached atomistically.
We observe that when the NRs are placed 6 \AA\ above the graphite surface, methane is blocked out, while CO$_2$ molecules  can  diffuse and be collected in between the NRs and the graphite surface. Consequently, the selectivity of \CO2 is extremely high. 
We also observe that the  initial rate of adsorption of  CO$_2$ is much higher than \ch4.  Overall we show that   the filter can be optimized by controlling the gap between NRs and the NRs-graphite separation.
\end{abstract}

\section{Introduction}
\label{intro}

Separating CO$_2$ from CH$_4$ is critical in industrial applications,
transportation, and usage of methane \cite{Bastin2008}.  Researchers
have intensively investigated the  adsorption of \co2/\ch4 mixtures on
several materials like MOFs, mesoporous carbon, activated carbon,
silicalite, C$_{186}$ schwarzite, and nanoporous carbon experimentally
or theoretically
\cite{Bastin2008,Babarao2007,Gatica2016,Heuchel1999,Liu2012,Sidi2016,Sidi2013,Palmer2011,Peng2009}.

Gas separation by adsorption can be accomplished by three  physical mechanisms: equilibria, kinetics, and steric effects. Equilibrium mechanisms rely on the strength of attraction between gas molecules and the substrate, while kinetic mechanisms involve the differences in the adsorption and transport rates of  gas on and through the  substrate \cite{Gatica2016}. Steric mechanisms, on the other hand, depend on the incompatibility between the size or shape of the adsorbate gas molecules and the pores of the substrate.
 For instance, since CO$_2$ is typically found in a mixture with gases
 of similar size but different shapes (as CH$_4$ and N$_2$), steric
 separation may be feasible. Also, the force of some substrates is
 stronger to CO$_2$ than methane. For example, in this study, we found
 that the energy of interaction of CO$_2$ with graphite is 35$\%$
 stronger than CH$_4$ with graphite. As a result, the equilibrium
 mechanism may also present an adequate strategy.

The ability of a substrate to separate gases by adsorption is measured by the selectivity. In a binary mixture of components $i$ and $j$, the selectivity is defined as,
 \begin{equation}\label{S}
  \Sigma (i/j)=\frac{ {x_{i}}/ {x_{j}}}{{y_{i}}/ {y_{j}}} ,
  \end{equation}
 where $x_{i}$ and $y_{i}$ are the molar concentration of species $i$
 in the adsorbed phase and  the vapor phase respectively. A high value
 of $\Sigma (i/j)$  indicates that the fraction of species $i$ in the adsorbed phase is large compared to the fraction in the vapor phase, meaning that $i$ is favorably adsorbed

In most studies of separation of \co2/\ch4 at room temperature, the
selectivities reported are smaller than  12.  One of the highest
selectivities found   has been achieved by  the group of Palmer et
al. \cite{Palmer2011}. They investigated different types of nanoporous
carbons  to separate  CO$_2$/CH$_4$ mixtures at ambient temperature
and pressure up to 10 MPa using Grand Canonical Monte Carlo (GCMC)
simulations.   They found that in the carbon slit pores, the
selectivity reaches 12.1 for a mixture with 25$\%$ CH$_4$ and at a
pressure of 3 MPa. In a mix with 50$\%$ CH$_4$ and pressure 4 MPa, the
selectivity is 11, whereas in a mixture with 75$\%$ CH$_4$ and
pressure 6 MPa, the selectivity is 9.2. 

Heuchel et al. \cite{Heuchel1999} obtained $\Sigma$(CO$_2$/CH$_4$)
experimentally and theoretically on activated carbon A35/4 at 293
K. They found  selectivities in the range from 2.8 to 8.9 for various
concentrations and pressures.
 Chen et
al. \cite{chen2017}, studied the 
\co2/\ch4 selectivity in a MOF-505@GO composite finding a value of 8.6 at
298 K and 100 kPa. In a recent article, Szcz\c{e}\'{s}niak et
al. \cite{SZCZESNIAK2020109761} reported \co2/\ch4 selectivity of 6.3
nand 5.8 
in the Cu-BTC MOF  and Cu-BTC/GO10 respectively, at 298 K and 1 bar.
Wang et al. \cite{wang2020} investigated \co2/\ch4  separation in
a penta-graphene (PG) nanosheet. They found that the selectivity is high 
when an electric field of 0.040 a.u. is applied. The effect of the
electric field is to change the adsorbed \co2 from physisorption to
chemisorption while not affecting the methane adsorption. 


Other groups found more modest values for selectivity. Bastin et
al. \cite{Bastin2008}   examined a microporous MOF (MOF-508b) for the separation and removal of CO$_2$ from binary CO$_2$/CH$_4$, CO$_2$/N$_2$, and ternary CO$_2$/CH$_4$/N$_2$ mixtures by fixed-bed adsorption at temperatures 303 K, 323 K, and 343 K. At 303 K for binary or ternary mixtures, the adsorption isotherms indicate that MOF-508b is moderately efficient for the removal of CO$_2$. They reported that  $\Sigma$(CO$_2$/CH$_4$) and $\Sigma$(CO$_2$/N$_2$) are in the range from 3 to 6 while decreasing with increasing   temperature.

 Liu et al. \cite{Liu2012} investigated the adsorption and the selectivity of CO$_2$ from CO$_2$/CH$_4$, CO$_2$/N$_2$ gas mixtures in oxygen-containing functionalized graphitic slit pores at a temperature of 298 K and pressures up to 300 bar using GCMC simulations. In their results,   $\Sigma$(CO$_2$/CH$_4$) was in the  range from 2 to 5.

 Babarao et al. \cite{Babarao2007}, predicted the value of $\Sigma$(CO$_{2}$/CH$_{4}$)   at room temperature in  three different adsorbents: silicalite, C$_{168}$ schwarzite, and IRMOF-1. By using GCMC, they found that the selectivity of CO$_2$/CH$_4$ is in the range from  2.0 to 3.2, 2.0 to 2.6, and 3.0 to 5.0 for IRMOF-1, silicalite, and the C$_{168}$ schwarzite, respectively.    Peng et al. \cite{Peng2009} investigated $\Sigma$(CO$_2$/CH$_4$) on ordered mesoporous carbon CMK-1 using GCMC  at T= 308 K and for different pressures and concentrations. Their results show  that the highest selectivity of CO$_2$ is 3.55 when P=7.0 MPa  and y$_{CO_2}$= 0.2.

Other groups evaluated the selectivity at low temperatures, finding extremely high  values. For instance, Gatica et al. \cite{Gatica2016}, reported a numerical study of adsorption of CO$_2$/CH$_4$ on carbon nanohorns, finding selectivities growing from 6 to 25  for a  temperature drop from  193 K  to   143 K. Maiga et al. \cite{Sidi2013,Sidi2016}, estimated the selectivity of a CO$_2$/CH$_4$ binary mixture in MOFs and graphene using GCMC and the ideal adsorbed solution theory (IAST). For MOFs, they predicted that the selectivity increased from 2 to  250 for temperatures dropping from 300 K to 120 K.  For graphene at 40 K, the selectivity estimated by GCMC-IAST  reaches the extremely high value of $6.24 \times 10^6$.  \cite{Sidi2016}

\begin{figure} []
 \includegraphics[width=8.5cm]{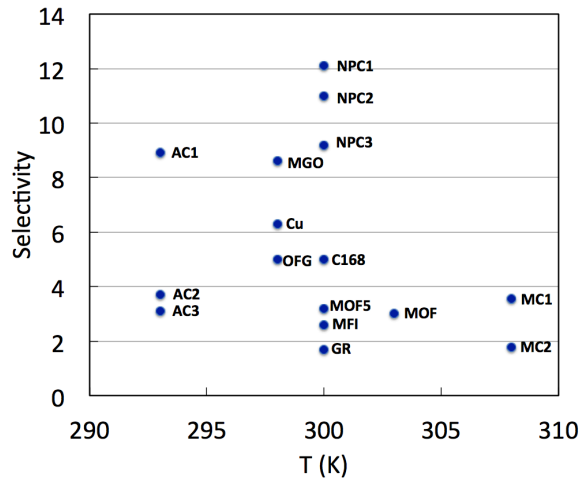}
\caption{ 
Selectivity for CO$_2$/CH$_4$  near room temperature using different
substrates: C$_{168}$, MOF5 and MFI from Ref. \cite{Babarao2007};
MOF-508b (MOF)  \cite{Bastin2008}; Activated Carbon at concentrations
of \co2  90$\%$ (AC1), 50$\%$ (AC2), and 20$\%$ (AC3),
respectively\cite{Heuchel1999}; Oxygen-containing functionalized
graphitic slit pores (OFG) \cite{Liu2012}; graphite (GR), from this work;
nanoporous carbon at 25$\%$  concentration of CH$_4$ and pressure 3
MPa (NPC1), 50$\%$ - 4 MPa (NPC2), and 75$\%$- 6MPa (NPC3)
\cite{Palmer2011}; mesoporous carbon CMK-1 at 20$ \%$ concentration of
CO$_{2}$ and pressure  7.0 MPa (MC1), and 50$\%$-4.0 MPa (MC2)
\cite{Peng2009}; MOF-505@GO at 100kPa (MGO)  \cite{chen2017}; Cu-BTC at 1 bar
(Cu). \cite{SZCZESNIAK2020109761} 
   }
 \label{fig: selectivity}
 \end{figure}

For practical reasons, it is vital to optimizing substrates capable of  separation by adsorption at room temperature. 
In Fig. \ref{fig: selectivity}, we summarize the values of selectivities near 300 K reported by   many groups.

We propose a simple  substrate to reach   high selectivity; it  consists of   graphene NRs  placed over  graphite.  The NRs are assembled parallel to each other and  the graphite surface.
 By tuning  the separation between them, the edge style,  and the
 distance between  the NRs and graphite, we reach  our objective.

 Graphene NRs  are quasi-one-dimensional carbon structures that can be
 obtained by cutting a graphene sheet into strips of a few nanometers
 width \cite{Saroka2014}.  Graphene is itself   a unique material with
 a high specific surface area of $\sim$ 2600 m$^{2}$/g,   which is
 obtained from graphite through oxidation, exfoliation, and reduction
 \cite{Gadipelli2014}.   Ever since the experimental isolation of
 graphene in 2004, significant research efforts have been focused on
 investigating the electronic and transport properties of its
 NRs. Several techniques have been developed  to fabricate
 them. These include electron beam lithography and etching, chemical
 synthesis, and unzipping of carbon nanotubes \cite{Orlof2013}. The
 properties of graphene nano-ribbons can be tuned from metallic to
 semiconducting through changing the widths and the edge styles
 \cite{Yin2013}.  Paulla et al. \cite{paulla2013}  investigated
 graphene NRs for the sensing of carbon oxides. In their ab initio
 study, they found that pristine NRs may not be suitable for
 electrochemical sensing of carbon oxides with low concentration at
 room temperature; they suggest that adding an electric field can
 generate detectable coverages. 

Although, in this work we do not address the technical aspects of
 keeping the graphene NRs fixed in place, we estimate that  this can
 be accomplished by inserting spacers in between the NRs and the
 graphite surface, resembling
 the structure of 
 MOFs or pillared graphene \cite{pillaredgraphene}.

 \section{Methods}
 
  In this work, we compute, by MD simulations,  the adsorption of a vapor mixture of \co2 and \ch4 on the NRs/graphite substrate. We represent CH$_4$ as a neutral spherical molecule with zero electric dipole and quadrupole moments.   The CH$_4$-CH$_4$ intermolecular interaction energy is modeled as a Lennard Jones (LJ) potential. 
The LJ potential  is given by, 
 \begin{equation}\label{lj}
 U_{LJ}=  4 \epsilon 
  \Big[ \Big( \frac{\sigma}{r} \Big ) ^{12} 
 - \Big(  \frac{\sigma}{r} \Big ) ^{6}  \Big],    
  \end{equation} 
where $\epsilon$ and $\sigma$  are the energy and size LJ parameters, respectively, and $r$ is the distance between atoms. The LJ parameters for \ch4  are $\epsilon$ = 148.0 K and $\sigma$ = 3.7 \AA\, adopted from  Ref. \cite{Babarao2007}.  
We describe the CO$_2$ molecule as a linear rigid body with three LJ sites and three partial charges placed on each atom. The carbon atom has a positive charge $q_{C}= 0.576 e$, and the oxygen atoms have a negative charge $q_{O}=- 0.288 e$. The bond length is $b$ =1.18 \AA \cite{Babarao2007}. The CO$_2$-CO$_2$ intermolecular interaction energy is  computed as a combination of the  LJ and Coulomb potentials between partial charges.
We adopt the LJ parameters from Ref. \cite{Babarao2007} (see table \ref{tab:LJ1 parameters}). 

\begin{table}[]
    \centering
 \caption{Lennard Jones Parameters and Partial Charges. }
    \begin{tabular}{lllll} \hline
    Adsorbate & $\epsilon$(K) & $\sigma $(\AA) & $q$($e$) \\ \hline
    CH$_4$ & 148.0 & 3.7 & 0      \\ \hline
    C in CO$_2$ & 29.70 & 2.8 &+0.576         \\ \hline
    O in CO$_2$ & 83.00& 3.0 & -0.288              \\ \hline
    C in Graphene &28.00 & 3.4& 0             \\ \hline
    \end{tabular}
    \label{tab:LJ1 parameters}
    \end{table} 

 For different species, the cross parameters are calculated by using the Lorentz-Bertholet combination rules\cite{Bruch1997},  
 \begin{eqnarray}
   \label{semi}
 \sigma_{ij} &= \frac{\sigma_{ii} + \sigma_{jj}}{2} \\
 \epsilon_{ij} &= \sqrt{\epsilon_{ii} \epsilon_{jj}}\label{semi2}
\end{eqnarray}

In our simulations, we treat graphite as a  continuous matter. The interaction energy between an adsorbate atom and the graphite substrate is evaluated by  the Steele-10-4-3 potential \cite{Steele1974} given by,

\begin{eqnarray}
\label{H}
\nonumber   U_{Steele}(z) & = & 2\pi \epsilon\rho \sigma^{2}\Delta \bigg[ \frac{2}{5}\left 
( \frac{\sigma}{z}\right )^{10}-\left ( \frac{\sigma}{z} \right )^{4}\\
&-& \frac{\sigma^{4}}{3\Delta \left ( z+0.61\Delta  \right )^{3}} \bigg],
\end{eqnarray}
%
 where $\epsilon$ and $\sigma$ are the LJ parameters and  $z$ is the distance between the adatom and the graphite surface; the number of carbon atoms per unit volume in graphite is $\rho$ = 0.114 \AA$^{-3}$, and the separation distance between the layers of graphitic carbon is $\Delta$ = 3.35 \AA.

 We describe the interaction between the NRs  and the adsorbates by a sum of LJ potentials.  The NRs  are kept rigid and fixed during the simulations.

We run MD simulations at constant temperature  by using the Nose-Hoover thermostat, which is based on the extended Lagrangian method.
 We run our simulations with the LAMMPS program, which stands for (Large-scale Atomic/ Molecular Massively Parallel Simulator). \cite{Plimpton1995,lammps}.

\begin{figure}[]
 \centering
 \includegraphics[width=8.4cm]{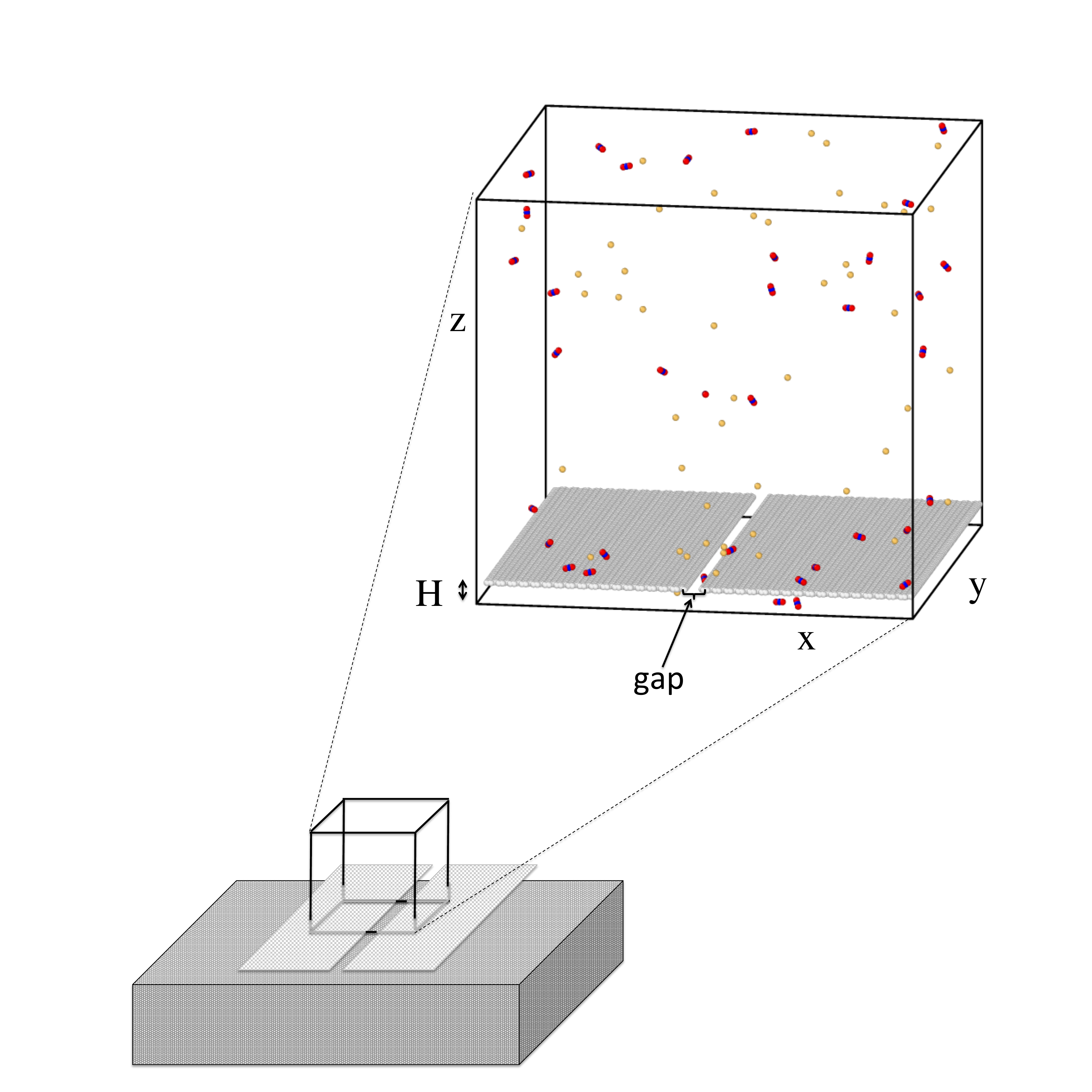}
\caption{Schematic picture of the nanoribbons over graphite and zoom of the simulation box.  }
 \label{fig: smbox}
 \end{figure}

\begin{figure}[]
\centering
 \includegraphics[width=7.5cm]{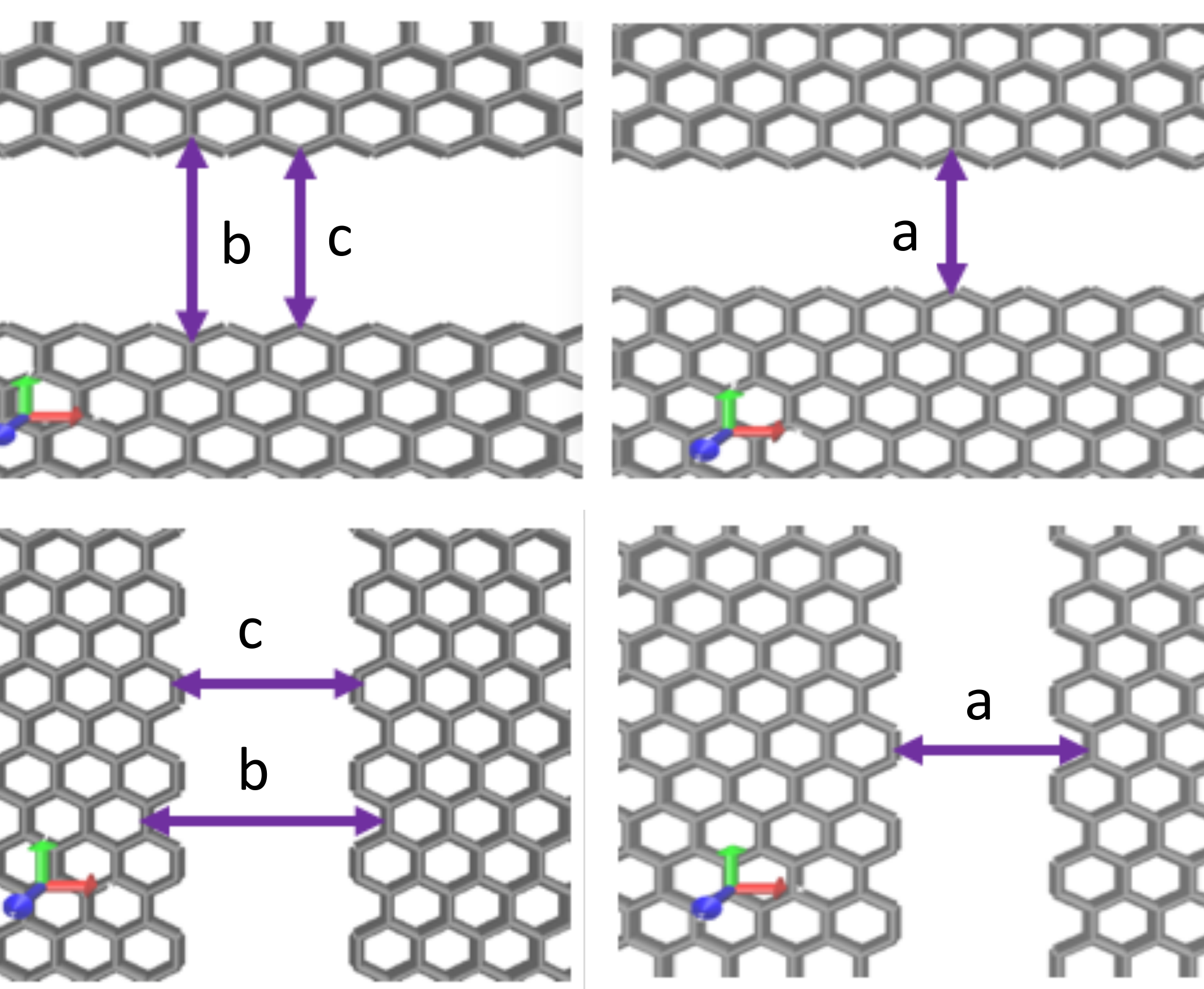}
    \caption{Illustration of the edges of the graphene nanoribbons with styles zigzag-VV (top left), zigzag-VB (top right),  armchair-VV (bottom left), armchair VB (bottom right). } 
\label{GNRs}
\end{figure}

  We set the simulation box with 160 \AA\ side lengths and  periodic
  boundaries in XY directions. The box contains the right and left
  halves of two nanoribbons. The NRs  are approximately 15-nm wide and
  infinitely long due to the periodic boundary conditions  (see
  Fig. \ref{fig: smbox}).  The top wall of the box  is reflective, and the graphite surface is at the bottom of the cell. To create the NRs we  place a graphene layer  above the graphite substrate and delete the carbon atoms within a selected narrow region. By cutting the graphene layer in the Y direction, we obtain zigzag NRs (ZRs), whereas cutting the graphene layer in the X direction, gives armchair NRs (ARs). By manipulating the region where the carbon atoms are deleted, we get two different styles of edges, as shown in Fig.  \ref{GNRs}. The figure   represents the ZRs in the upper row and ARs in the lower row with two different edge styles:   VV and  VB. In style VV, the vertexes are facing vertexes, and the bays are facing bays. For style VB where the vertexes are facing the bays, the gap opening is characterized by  parameter a. For style VV, there are two   parameters: b and c, which are the distance from bay to bay and vertex to vertex, respectively. To fine-tune the gap separation, we translate the NRs horizontally by tenths of angstroms.

The simulations are composed of two stages. First, we run a 1-million-steps MD of 50 \CO2 and 50 \CH4 molecules in the plain simulation box, to achieve a mixed vapor in equilibrium at 300 K.  In the second stage, we include the graphite substrate and the NRs in the simulation box and run MD for 5 million fs-timesteps.
Molecules are attracted to graphene and pass through the slit forming
layers on the graphite surface and the bottom of the   ribbons.

The usual definition of selectivity given in Eq. \ref{S} assumes that
the system is in equilibrium with a vapor.  That is not the case in
MD simulations, where the process   described is rather dynamic. Some
molecules leave the vapor to occupy the region next to the substrate,
while the total number of molecules in the box remains constant.
Depending on the amount adsorbed, the molecules remaining in the vapor
phase are only a few or even zero.   For this reason, we redefine the
selectivity as

 \begin{equation}\label{S2}
S= \frac{x_{\textrm{CO}_{2}}/x_{\textrm{CH}_{4}}}{y_{\textrm{CO}_{2}}/y_{\textrm{CH}_{4}}}  
\end{equation}
 where $x_{i}$ is the  concentration of species $i$  in the adsorbed phase,  and $y_{i}$  is  the concentration in the simulation box. We compute   $x_{i}$   by counting   the number of molecules in a region of thickness 5-\AA\ above the graphite surface.  To calculate the selectivity, we use the average of $x_{\textrm{CO}_2}$ and $x_{\textrm{CH}_4}$ over the last million steps (last nanosecond).

To asses the kinetics of the process we define the relative rate of adsorption $R$ as,
   \begin{equation}\label{R}
   R=\frac{r_{\textrm{CO}_{2}}}{r_{\textrm{CH}_{4}}},
   \end{equation}
 where $r_{\textrm{CO}_2}$ and $r_{\textrm{CH}_4}$ are the initial rate of   adsorption  for CO$_2$ and CH$_4$,  respectively. The rates $r_i$  are calculated by a linear fit of the fraction adsorbed vs. time on the first nanosecond.

\section{Results}

We studied  various combinations of  gap openings,   NRs-graphite distances  (from 6 \AA\ to 14 \AA), and edge-styles (zigzag and armchair,  VV and VB).   
In Table \ref{results}, we present the values of the selectivity and
relative rate of adsorption for selected NRs with $S>2$ or $R>2$.    In the cases where the adsorption of \ch4 is zero during the first nanosecond interval, we report $R \rightarrow \infty$. Also, in the case of  $x_{CH4}=0$ during   the last nanosecond interval, the selectivity is infinite.

\begin{figure} []
 \centering
 \includegraphics[width=8.4cm]{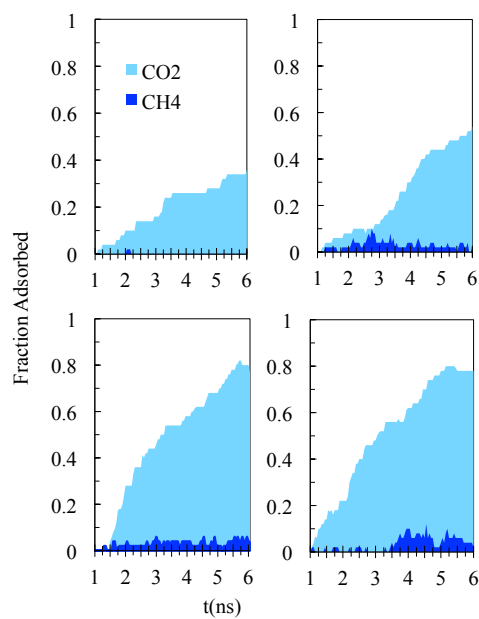}
\caption{Fraction adsorbed on graphite $x_{CO2}$  (lightblue) and $x_{CH4}$ (blue) on the graphite surface. The graphene ribbons are placed at a distance H= 6 \AA\ above the graphite.   The edge styles are  
armchair VB with parameter  a = 7.4 \AA\ (top left),  
armchair VV with b = 9.8 \AA\ and c = 7.4 \AA\ (top right),
armchair VV with b = 14.8 \AA\ and c = 12.3 \AA\ (bottom left) 
and zigzag VB with   a  = 12.4 \AA\ (bottom right).
} 
 \label{fig:blocking}
\end{figure}

In Fig.  \ref{fig:blocking} we show the results for the  top-four selectivities. In the figure,  we plot  the adsorption of  \co2 and \ch4  on the graphite surface as a function of time. In the four  cases, the graphene NRs are located  at a distance H  = 6 \AA\  above graphite.  
The maximum selectivity  was  obtained for   VB-edged ARs, with a vertex-bay distance of 7.4 \AA\ (see Fig. \ref{fig:blocking}, top left).
 In this case, \ch4 is wholly blocked out giving   $S  \rightarrow \infty  $, and also due to zero \ch4-adsorption in the first nanosecond,   we obtain   $R \rightarrow \infty$. 
The second top value of selectivity is 22 (Fig. \ref{fig:blocking} top right), for   VV-edged ARs with parameters b  = 9.8 \AA\ and c = 7.4 \AA; the value of $R$ is 2.7.
The third  top value of selectivity is 21,  shown in Fig \ref{fig:blocking} (bottom left).  This case corresponds to    VV-ARs  with parameters b = 14.8 \AA\ and c = 12.3 \AA. 
The fourth  top value is $S=15$,  achieved with    VB-ZRs  with
parameter a = 12.4 \AA.   The last two cases also have   high values of $R$ (14 and 18).

\begin{figure} []
 \centering
 \includegraphics[width=8.4cm]{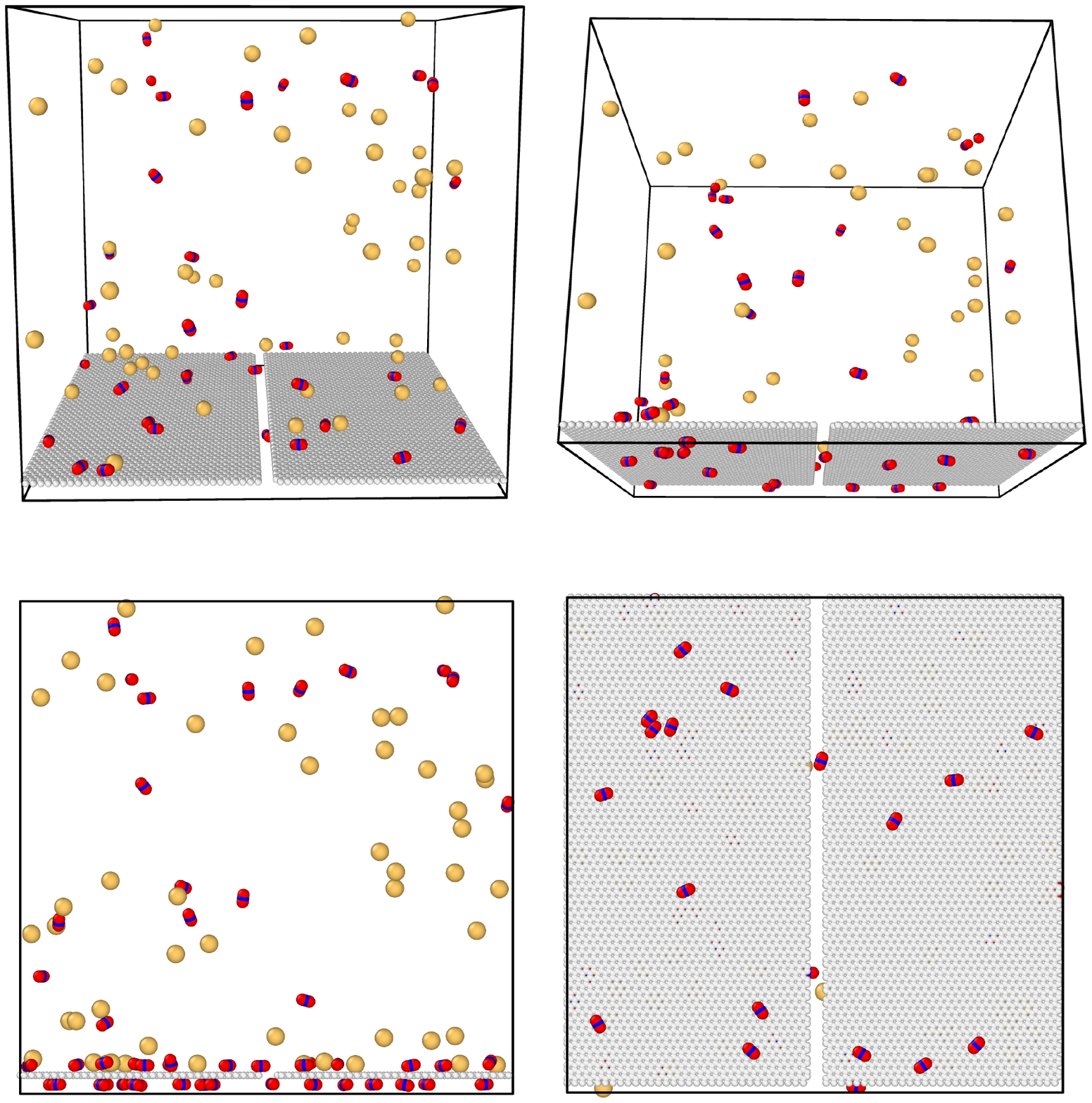}
 \caption{Snapshots of the simulation at t = 6 ns,  showing  \ch4 in yellow and \co2 in blue-red. The molecules are not in scale. The graphite surface is not shown for clarity.
The panels display two perspective views   (top row), a side view (bottom left) and a view from below the NRs (bottom right). The NRs' edges are armchair-VB with a 7.4-\AA\ gap, and H = 6 \AA.
} 
 \label{snapshot1}
 \end{figure}

\begin{figure} []
 \centering
 \includegraphics[width=8.4cm]{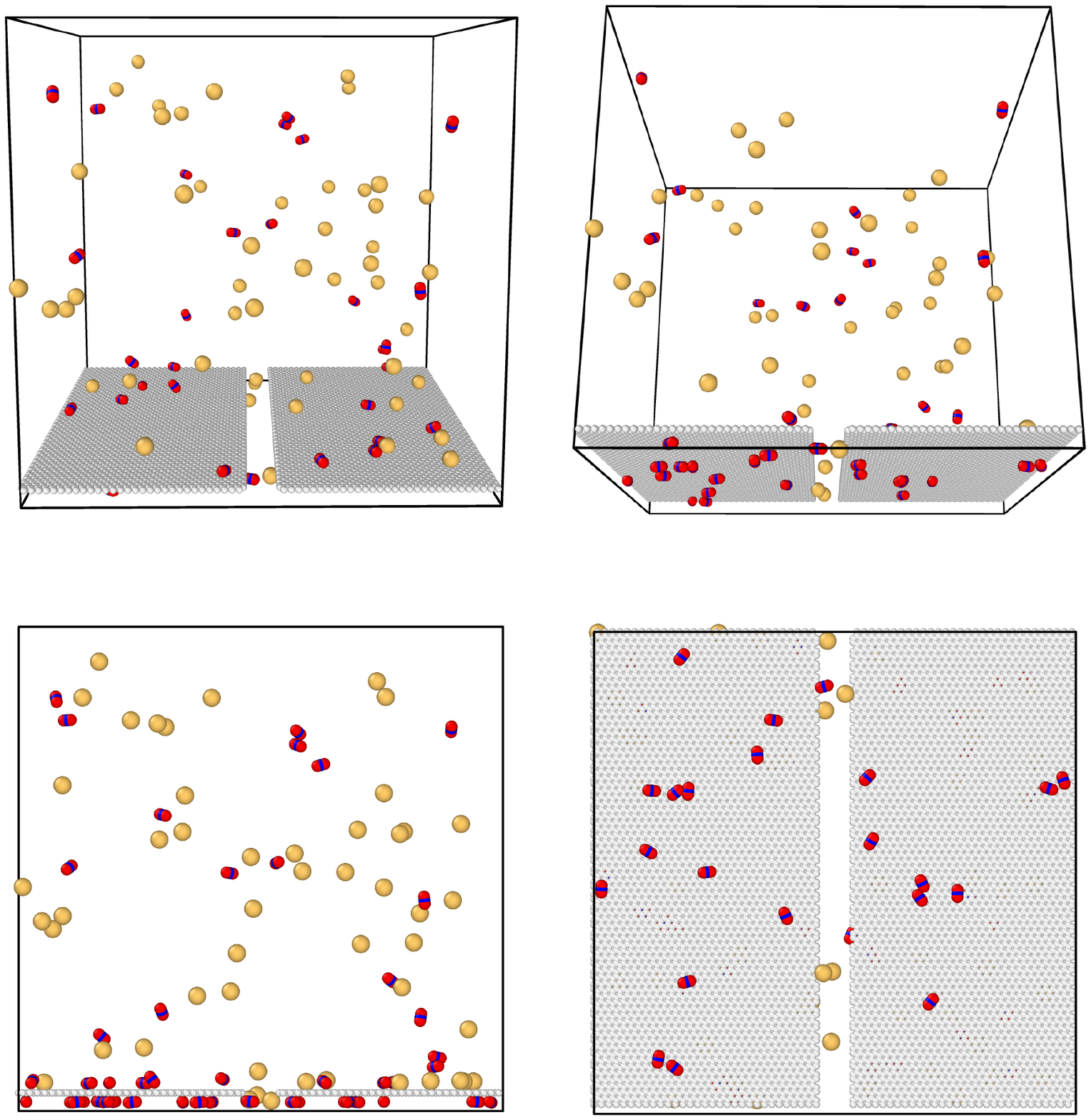}
\caption{Snapshots of the simulation at t = 6 ns,  showing  \ch4 in yellow and \co2 in blue-red. The molecules are not in scale. The graphite surface is not shown for clarity.
The panels display two perspective views   (top row), a side view (bottom left) and a view from below the NRs (bottom right). The NRs' edges are armchair-VV with a 14.8-\AA\ gap, and H = 6 \AA.
} 
 \label{snapshot2}
 \end{figure}

In all cases with H  = 6 \AA\, methane does not fit in the space
between the NRs and graphite. However, for wider gaps (second, third,
and fourth cases),  methane molecules are adsorbed on graphite at the
slit opening. To exemplify, we show snapshots of the simulation cell at
t = 6 ns in Figs. \ref{snapshot1} and \ref{snapshot2}. Both cases
correspond to H = 6 \AA\ with gaps 7.4 \AA\  (Fig. \ref{snapshot1})
and 14.8 \AA\ (Fig. \ref{snapshot2}). For the 14.8-\AA\ gap, we see in
Fig. \ref{snapshot2},  a few \ch4 molecules   adsorbed on graphite right  at  the slit.

As a reference for comparison, we simulated adsorption of the mixture on plain graphite,  obtaining  $S= 1.68$ and $R= 1.98$.  As it results, the selectivity and $R$ achieved  with  both ZRs and ARs are much higher than plain graphite.  

When the distance between the graphene and graphite is higher than 8 \AA\ and the gap opening  wider than 6.9 \AA\ in ZRs or 7.4 \AA\ on ARs, the   values of $S$ and $R$ are not better than plain graphite.  
On the other hand,  any  smaller gap opening yields zero adsorption, because the slit  then becomes too narrow for any molecule to pass. A distance H $<  6$ \AA\ results in zero adsorption as well, because the NRs-graphite separation becomes too thin to allow any of the molecules within. 

The structure of the molecules play a crucial role; only the linear CO$_2$  fits in the 6-\AA\ narrow space between the NRs and graphite  while  the spherical   CH$_4$ does not.    
 Moreover, the Van der Waals interaction between CO$_2$ and  graphite is stronger than   CH$_4$-graphite. In our simulations, we computed the average   adsorption energy between graphite and CO$_2$,  E$_{G-CO_{2}}$ = 3.76 Kcal/mol, while    E$_{G-CH_{4}}$ = 2.64 Kcal/mol. As a result, the selectivity for CO$_2$ is high due to a combination of energetic and steric effects.

The kinetics of adsorption is also greatly  improved. In plain graphite, the initial  uptake   of \co2 is just  twice as  fast as  for \ch4. 
In the NRs/graphite substrate, the   relative rate $R$  reaches  18 and 14   for ZRs and ARs, respectively.

 \section{Discussion}

Our findings show that  the  separation of a carbon dioxide and methane mixture on NRs-Graphite at room temperature  is  significantly  enhanced.  All the NRs edges tested in this work are adequate for filtering carbon dioxide.

The mechanisms involved in the performance of the filter combine
energetics, kinetics, and steric aspects of adsorption. The energetics
works because \co2 is more intensively attracted to the substrate; the
kinetic factor arises from the higher initial rate of adsorption of
\co2 even in non-blocking cases.
There is an interplay of adsorption and diffusion above and below the
graphene NRs and on graphite. \co2 adsorbs first  on top of the
graphene,  followed by \ch4. Then, \co2 diffuses through the gap and
forms  monolayer films on  the graphite surface and below the NRs. If
the distance between the NRs  and graphite is  6 \AA, \ch4 is
completely blocked  out  even with wide enough NRs gap openings.  
This is a consequence of the overlap of the repulsive cores of the
\ch4-graphene and \ch4-graphite interactions that results in a total
repulsion of methane.
In the non-blocking cases, \ch4 molecules would start adsorbing on graphite later than \co2.

The optimal parameters for the filter are a NRs-graphite separation of 6 \AA,   armchair-VB  edge style,  and  gap opening   of 7.4 \AA. However, slightly larger separations and openings and other edge styles are also effective.

As mentioned above, although we have not addressed the technical aspects of keeping the
graphene NRs fixed in place, we would estimate that  this can be
accomplished by inserting organic molecules as pillars resembling the
structure of pillared graphene \cite{pillaredgraphene}.
However, the optimal  separation of 6 \AA\ may be too small to be
achieved by a
pillared structure. In this case, we would envision a different
approach  similar to the graphite intercalation compounds \cite{GIC}.  Of course, the
technicality is far from the scope of our work. 
However we must say
that adding such pillars or spacers would affect the adsorption rates
but not the fact that methane is blocked out when the separation is 6 \AA\ wide.
Finally, the reader may argue that  the proposed substrate lacks the capacity
for 
storage. That is true, and we would like to emphasize
that our results pertain to the filtering  rather than the storage.
As such,  the filter should be combined with a supplementary substrate for
that purpose.

\begin{table}[h!]
  \begin{center}
    \caption{Selectivity and relative rate of adsorption for selected slit openings.}
    \label{results}
    \begin{tabular}{lllll} 
      \hline
      H({\AA}) & Style & Gap({\AA})  & $S$  & $R$ \\
      \hline
      6 &  A VB & 7.4       & $\infty$   &$\infty$ \\
      6 &  A VV & 9.8      & 22   &2.7\\
      6 &  A VV &14.8     &21  &14\\
      6 &  Z VB & 12.4    &15   & 18\\
      6 &  Z VV & 14.2    & 11   & 9.6\\
      6 &  Z VB & 6.9     &8.8   & 12\\
    12& A VB & 7.4       & 8.6 & 4.0\\
       12& Z VB &  6.9    &8.4 &14\\
      6 &  A VV & 12.3  &7.9   &11\\
        10& A  VB& 7.4   & 7.5  & 8.5\\
      14 &A VB& 7.4      & 7.0   &4.3\\ 
          6 &  Z VV & 9.8  &6.5   & 6.9\\
         10& Z  VB& 6.9  & 6.5  & 2.7\\
     14 & Z VB &  6.9    & 6.1& $\infty$ \\    
         8 &  Z  VB& 6.9   & 5.9 & 2.7\\
         8 &A VB & 7.4    & 5.5 &3.8\\
       14& Z  VB & 12.4    & 1.9 & 3.1\\
      14 & Z  VV& 9.8     & 1.8  & 3.3 \\

      \hline
    \end{tabular}
  \end{center}
\end{table}

\begin{acknowledgements}
This work used the Extreme Science and Engineering Discovery Environment (XSEDE) through allocation TG-DMR180036.
\end{acknowledgements}

%
\section*{Conflict of interest}

 The authors declare that they have no conflict of interest.

\bibliographystyle{spphys}
\bibliography{hindbib.bib}

\end{document}